\documentclass[twocolumn,preprintnumbers,superscriptaddress]{revtex4}
\usepackage[utf8]{inputenc}
\usepackage{amsmath}
\usepackage{dcolumn}
\usepackage{bm}
\usepackage{epsfig}
\usepackage{epstopdf}
\usepackage{graphicx}
\usepackage{amsfonts}
\usepackage{amssymb}
\usepackage{mathrsfs}
\usepackage{color}
\usepackage{multirow}
\usepackage{appendix}
\usepackage{natbib}
\usepackage{url}
\usepackage{doi}
\usepackage{soul, color, xcolor}
\soulregister{\cite}7 

\usepackage{hyperref}
\hypersetup{
    colorlinks=true,
    linkcolor=blue,
    citecolor=blue,
    urlcolor=blue
}

\newcommand{\APM}{Wuhan Institute of Physics and Mathematics, Innovation Academy of Precision Measurement Science and Technology, Chinese Academy of Sciences, Wuhan 430071, China}
\newcommand{\UCAS}{University of Chinese Academy of Sciences, Beijing 100049, China}
\newcommand{\GZII}{Guangzhou Institute of Industrial Intelligence, Guangzhou 511458, China}
\newcommand{\GZIIT}{Research Center for Quantum Precision Measurement, Guangzhou Institute of Industry Co., LTD, Guangzhou, 511458, China}

\begin{document}

	\setlength{\textfloatsep}{2pt}
	\setlength{\intextsep}{2pt}
	
	\title{Enhanced detection of electric-field signals via squeezing-induced stochastic resonance}
	\author{Ya-Qi Wei}
    \affiliation{Department of Mathematics, South China University of Technology, Guangzhou 510640, China}
    \affiliation{Laboratory of Quantum Science and Engineering, South China University of Technology, Guangzhou 510641, China}
    \affiliation{\APM}
    \author{Tai-Hao Cui}
    \affiliation{\APM}
    \affiliation{\UCAS}
    \author{Quan Yuan}
    \affiliation{\GZII}
    \affiliation{\APM}
    \author{Pei-Dong Li}
    \affiliation{\APM}
    \affiliation{\UCAS}
    \author{Yuan-Zhang Dong}
    \affiliation{\APM}
    \affiliation{\UCAS}
    \author{Zhuo-Zhu Wu}
    \affiliation{\APM}
    \affiliation{\UCAS}
    \author{Ji Li}
    \affiliation{\GZIIT}
    \author{Jia-Wei Wang}
    \affiliation{\APM}
    \affiliation{\UCAS}
    \author{Fei Zhou}
    \affiliation{\APM}
    \author{Ming-Xiao Li}
    \affiliation{Department of Mathematics, South China University of Technology, Guangzhou 510640, China}
    \affiliation{Laboratory of Quantum Science and Engineering, South China University of Technology, Guangzhou 510641, China}
    \author{Liang Chen}
    \email{liangchen@wipm.ac.cn}
    \affiliation{\APM}
    \author{Zhu-Jun Zheng}
    \email{zhengzj@scut.edu.cn}
    \affiliation{Department of Mathematics, South China University of Technology, Guangzhou 510640, China}
    \affiliation{Laboratory of Quantum Science and Engineering, South China University of Technology, Guangzhou 510641, China}
    \author{Mang Feng}
    \email{mangfeng@wipm.ac.cn}
    \affiliation{\APM}
    \affiliation{\GZIIT}
    \affiliation{Key Laboratory of Low-Dimensional Quantum Structures and Quantum Control of Ministry of Education,
                  Department of Physics and Synergetic Innovation Center for Quantum Effects and Applications, Hunan Normal University, Changsha 410081, China}
    \affiliation{Department of Physics, Zhejiang Normal University, Jinhua 321004, China}
	
	\begin{abstract}
		Stochastic resonance (SR) could amplify weak electric-field signals in nonlinear systems by means of the externally injected noises. Here we propose and experimentally demonstrate a modified SR method, termed squeezing-induced SR, implemented in the system involving a trapped ion behaving as a Duffing oscillator. We find that squeezing the phase noise of the  oscillator results in amplified fluctuation of the corresponding amplitude, which helps achieve the SR. Since no auxiliary noise source is needed, the squeezing-induced SR may enhance the signal-to-noise ratio by 4.28 $\pm$ 0.39 dB compared to the conventional noise-induced SR under identical conditions of the electric-field detection. This technique offers a promising approach for developing atomic ion sensors for detecting weak electric-field signals.

	\end{abstract}	
	\maketitle
	
	\section{Introduction}

Precision measurement of electric fields is a key means to reveal electromagnetic mechanisms in different research fields, from biological or biomedical studies \cite{Aslam2023} to particle sensing \cite{Yu2021} and even gravitational wave detection \cite{Yu2018}. Various devices for electric-field detection have been developed, including nitrogen vacancy centers \cite{Dolde2011}, semiconductor superlattices \cite{Shao2018},  nanowire cantilevers \cite{Braakman2014, Badzey2005}, Rydberg atoms \cite{Sedlacek2012, Jing2020, Holloway2022, Liu2022, Wu2024}, and trapped ions \cite{Maiwald2009, Biercuk2010, Liu2021, Kevin2021, Wei2022, Deng2023, Wei2023, Bonus2025, Wu2025, Blums2018}. In particular, trapped-ion systems, with their intrinsic sensitivity to external electric fields, well-controlled manipulation, and efficient collection of fluorescence signals, are excellent candidates for highly precise detection of electric-field signals.

In actual physical measurements, since noise from the environment is inevitable \cite{Brownnutt2015, Gardiner2000}, reducing its detrimental effects is essential to improving measurement precision \cite{Kuo1999, Khodjasteh2010, Suter2016}. However, in nonlinear systems, in contrast to the conventional notion that noise is inherently harmful, noise can sometimes be helpful for enhancing the systems' responses to weak periodic signals, revealing the constructive role of noise in information processing. A typical example is stochastic resonance (SR), which counterintuitively demonstrates the amplification of weak signals with strong background noise using the input noise itself \cite{Gammaitoni1998, Wellens2004, McNamara1988, Hibbs1995, Yuan2024}. This behavior is similar to the motion of a fictive particle in a double-well potential periodically modulated in amplitude by the signal under the influence of noise.

However, in a trapped-ion platform, the commonly used approach of injecting artificial noise is typically realized by applying controlled voltage noise to the trap electrodes, which directly increases the electric-field noise experienced by the ion. Such additional electric-field noise is known to induce excess motional heating of the trapped ions \cite{Brownnutt2015}, which brings extra instability to the confinement of ions and adds practical complexity to conventional noise-induced SR implementations.
To better extract weak signals from the strong background noise, we propose and experimentally demonstrate, in the present work, a modified SR method by replacing artificially introduced noise with squeezing signals. The key idea of the method, called squeezing-induced SR, is to redistribute the size of the thermal noise of the oscillator system in phase space, i.e., to amplify the noise in a certain dimension (e.g., the amplitude) of the system by squeezing the noise in the orthogonal dimension (e.g., the phase). As no extra noise is necessarily input, squeezing-induced SR works better for the sensitivity of electric-field detection than the conventional noise-induced SR under the same conditions.

Our experimental system is a cold trapped $^{40}$Ca$^{+}$ ion confined in a surface electrode trap (SET), as sketched in  Fig. 1. The ion is first Doppler cooled and then driven by a near-resonant electric field to form a forced nonlinear Duffing oscillator. By introducing a squeezing signal, we achieve detection of the weak electric field of the order of magnitude of $\mu$V/cm, which is an improvement of 4.28 $\pm$ 0.39 dB in signal-to-noise ratio (SNR) compared with counterparts using traditional noise-induced SR sensors, as elucidated later. Our work leverages the characteristic of squeezing into the weak-signal detection, which opens an avenue for developing nonlinear atomic ion SR sensors.
	
	\section{Experimental system and scheme}
		\begin{figure*}[htbp]
		\centering
		\includegraphics[width=16.6 cm,height=12.6 cm]{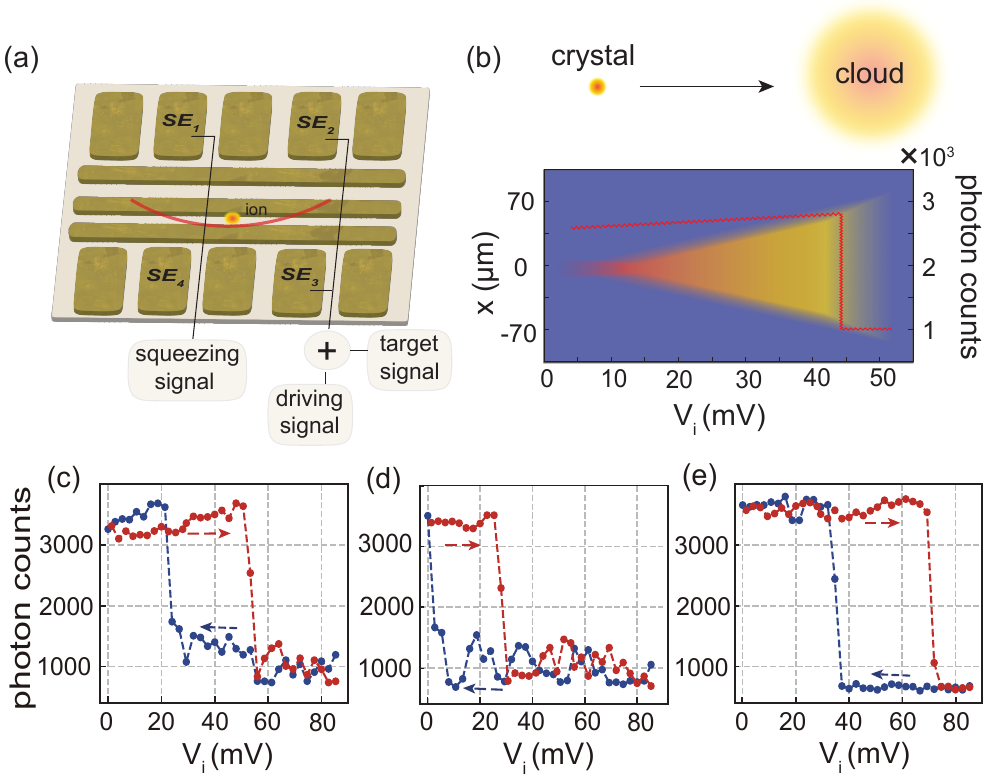}
		\caption{ Schematic diagram of the squeezing-induced SR and the experimental setup. (a) Top view of the SET, where the squeezing signal is applied to the $SE_{1}$ electrode, while the driven voltage amplitude $V_{i}$ and the target signal $E_{s}$ are both applied to $SE_{2}$ and $SE_{3}$ electrodes. (b) The driven electric field makes the trapped ion experience a transition from the crystalline state (with small distribution along the x-axis) to the cloud state (with large distribution along the x-axis), during which the photon counts drop suddenly. The orange area represents the vibrational amplitude of the ion, and the red line represents the magnitude of the photon counts. (c–e) Hysteresis loops monitored by photon counts with respect to the driven electric field, where each data point is measured with a 200-ms integral time. The sweeping directions are labeled with arrows. (c) No squeezing signal applied; (d) the squeezing signal with a gain of 0.98 and phase set to be $\pi/2$; (e) the squeezing signal with gain of 0.98 and phase set to be 0.}
		\label{Fig1}
	\end{figure*}
	
In our experiment, the SET is a 500-$\mu$m-scale planar trap, as introduced previously \cite{House2008,Wan2014,Liu2020}, whose secular frequencies are, respectively, $\omega_{z}/2\pi$ = 163.52 $\pm$ 0.01 kHz, $\omega_{x}/2\pi$ = 547.18 $\pm$ 0.10 kHz, and $\omega_{y}/2\pi$ = 797.50 $\pm$ 0.10 kHz. The single $^{40}$Ca$^{+}$ ion, at 800 $\mu$m above the surface of the SET, behaves as an atomic probe to detect the externally applied electric field; see Fig. \ref{Fig1}(a).  The first step toward SR is to generate a nonlinear Duffing oscillator. To this end, we apply a near-resonant electric field to the electrodes $SE_{2}$ and $SE_{3}$, which drives the trapped ion to oscillate with increasing amplitude and experience a transition from the crystalline state to the cloud state, as shown schematically in Fig. \ref{Fig1}(b). During this process, we observe the hysteresis loops \cite{blumel1988}, monitored by photon counts, by sweeping the voltage amplitude of the driven electric field back and forth; see Fig. \ref{Fig1}(c,d,e).

The target signal is a weak electric field of amplitude $E_{s}$ and frequency $\omega_{s}$, applied to electrodes $SE_{2}$ and $SE_{3}$ as an amplitude modulation of the strong driven electric field (with the driven voltage amplitude $V_{i}$ and frequency $\omega_{i}$ ( $\omega_{i} \approx \omega_{x}$)). This modulation perturbs the ion’s oscillation amplitude and synchronizes its dynamics with the weak signal. In addition, a squeezing signal of frequency $\omega_{sq}$ (with $\omega_{sq} = 2 \omega_{i} \approx 2 \omega_{x}$) is applied to electrode $SE_{1}$ to enhance detection of the weak signal $E_{s}$. Physically, the squeezing redistributes the thermal noise in phase space (Appendix A): by transferring noise from phase to amplitude, it enhances noise-assisted bistability switching near $\omega_{s}$, thus improving the weak-signal SNR.

\begin{figure*}[htbp]
		\centering
		\includegraphics[width=16.6 cm,height=12.6 cm]{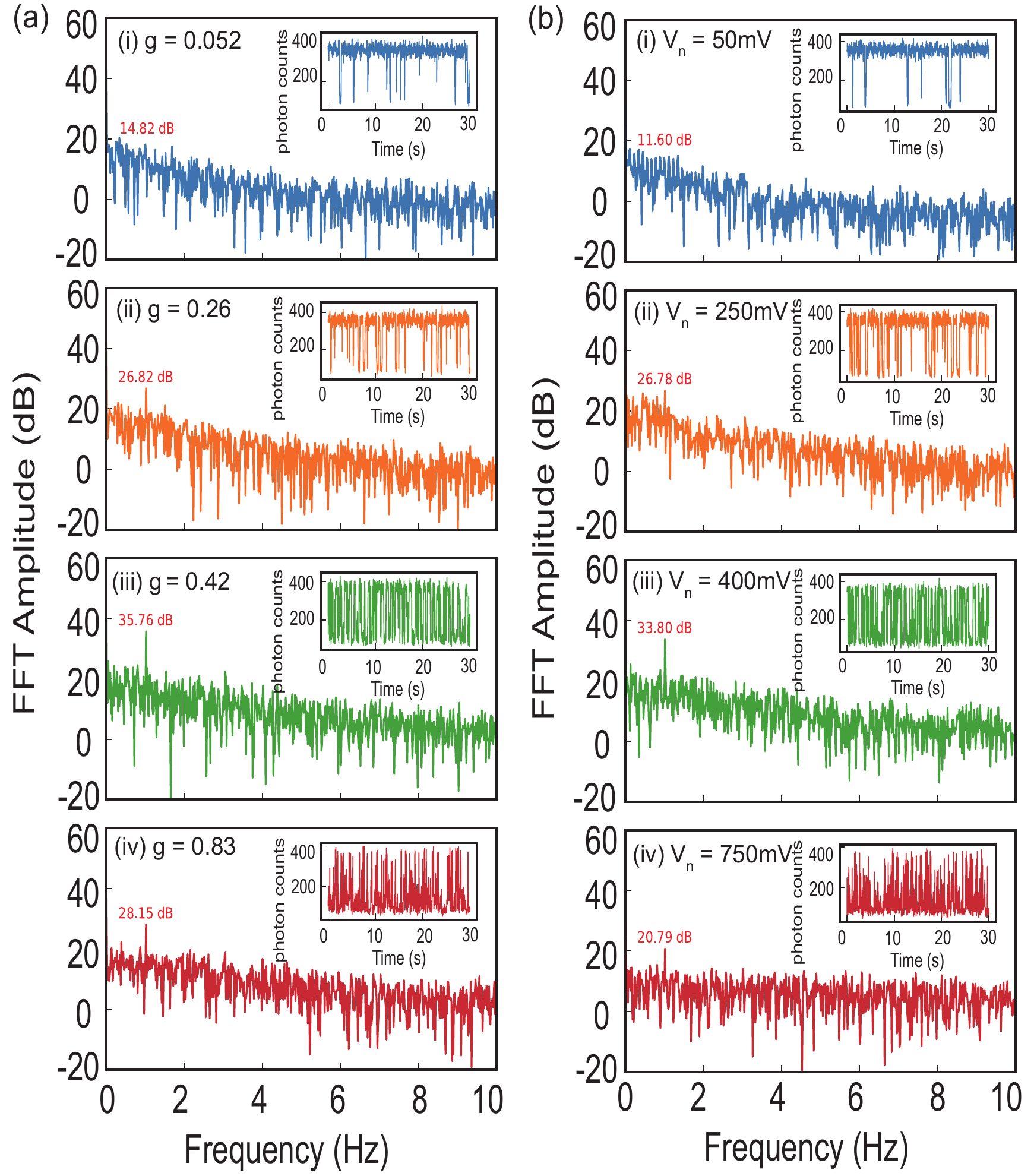}
		\caption{Comparison of squeezing-induced and noise-induced SRs. (a) Squeezing-induced SR reflected in the FFT spectra, in the squeezing signal gain increasing from top to bottom. (b) Conventional noise-induced SR reflected by the FFT spectra with the white-noise strength increasing from top to bottom, where the frequency bandwidth of the noise signal is 1 kHz. In both panels, the insets show the time-domain optical readouts (detected by PMT) with each data point measured by a 20-ms integral time.
        The target signal has a strength of $E_{s}$ = 2357 $\mu$V/cm and the frequency of 1 Hz. The driven voltage amplitude is $V_{i}=$47 mV and the frequency of $\omega_{i}/2\pi$ = 547.18 kHz. When the squeezing gain $g$ (or noise strength $V_{n}$) exceeds the threshold, stable signal peaks emerge. In this regime, the bistable state switch becomes synchronized with the target signal (see Appendix D for more details).}
		\label{Fig2}
\end{figure*}

For a trapped ion with mass $m$ and oscillating frequency $\omega_{x}$ under the friction coefficient $\gamma$ \cite{Vahala2009}, the dynamics can be described in a good approximation as a forced nonlinear Duffing oscillator,
	\begin{equation}
	\begin{split}
		\ddot{x}(t) + \gamma \dot{x}(t) + \left[ \omega_{x}^2 + g \gamma \omega_{x} \sin(\omega_{sq} t + 2\phi) \right] x(t) + \beta x^3 \\
		= \frac{F_{i} \left[ 1 + q E_{s}  \right] \cos(\omega_{i} t) + f_{n}(t)}{m},
	\end{split}
		\label{eq1}
	\end{equation}
where ${x}(t)$ denotes the displacement of the ion, and $\beta$ is the nonlinear coefficient. In our experiment, the electric fields of the driven signal and the target signal take the forms $F_{i}$ and $qE_{s}$, respectively, with $F_{i}$ generated by an ac signal $V_{i}(t) = V_{i}\cos(\omega_{i} t)$. Here, $g$ is the gain factor originating from the squeezing signal (see Appendix A for details), which modifies the trapping potential, and $\phi$ is the relative phase with respect to the driven electric field. Additionally $f_{n}(t)$ is the stochastic force related to the temperature $T$ that obeys the following ensemble average relations: $\langle f_{n}(t) \rangle$ = 0 and $\langle f_{n}(t)f_{n}(\tau) \rangle = 2 \gamma k_{b} T \delta(t - \tau)$, where $k_{b}$ is the Boltzmann constant, $T=10$ mK is the ion's effective temperature, and $\delta$ denotes the Dirac delta function.
	
Figure \ref{Fig1}(c-e) present hysteresis loops of the photon counts with respect to the strength of the driven electric field, reflecting steady bistable states. To understand the action of the squeezing signal, we first consider the situation without the squeezing signal applied, as presented in Fig. \ref{Fig1}(c). Then we apply the squeezing signal with the same gain $g = 0.98$ but different phases, where $\phi=\pi/2$ ($\phi=0$) indicates the squeezing of phase noise (amplitude noise) of the system. For $\phi=\pi/2$, the amplitudes of both the noise and the ion are amplified, leading to the occurrence of the bistable state at smaller $V_{i}$, see Fig. \ref{Fig1}(d). In contrast,  Fig. \ref{Fig1}(e) demonstrates the amplification of the phase noise along with the squeezing of the amplitude noise. In this case, a larger $V_{i}$ is required to reach the bistable state. Since a weaker driven field helps detect smaller target signals, we carry out the present experiment following the case in Fig. \ref{Fig1}(d).

\section{Experimental Results and Discussion}
\subsection{Demonstration of squeezing-induced versus noise-induced stochastic resonance}
Experimentally, the following steps are taken for both the squeezing-induced and conventional noise-induced SRs. We first apply the driven voltage signal $V_{i}$ and the squeezing (noise) signal $g$ ($V_{n}$), simultaneously, to the electrodes of the trap. By sweeping the amplitude of the driven electric field back and forth, we may acquire appropriate values for the characteristic parameters of hysteresis loops. Then we apply the target signal $E_{s}$ on the same trap electrode as $V_{i}$ applied, which behaves as the amplitude modulation of $V_{i}$.

To justify the squeezing-induced SR, we compare it with the conventional noise-induced SR under identical experimental conditions, such as drive amplitude, detuning, operating-point/bistability barrier depth, and total measurement time. The only difference is the switch-activation mechanism; namely, an injected Gaussian white-noise drive with strength $V_{n}$ and bandwidth $B_{noise}$ for noise-induced SR versus the squeezing signal drive for squeezing-induced SR. Based on the monitored photon counts, as shown in the insets of Fig. \ref{Fig2}, we present fast Fourier transform (FFT) spectra for different values of squeezing gain and noise strength. The signal recorded by the photomultiplier tube (PMT) is the spontaneous emission fluorescence from the ion. Concretely, we collect and detect resonance fluorescence due to the $4P_{1/2}\!\rightarrow\!4S_{1/2}$ transition of $^{40}$Ca$^{+}$, driven by a red-detuned 397-nm cooling laser (as described in Ref. \cite{Wei2023}). The 397-nm cooling beam is always on throughout the experimental implementation, providing a continuous readout via fluorescence; each panel corresponds to a 60-s record of the PMT output acquired at a sampling rate of 50 Hz. For clarity, only the 0–10 Hz portion of the FFT spectrum is shown; the time-domain inset displays a 30-s segment for illustration. As seen in Fig. \ref{Fig2}, when the strength of the squeezing signal or noise is weak, the ion remains in one of the two stable states. With the increase in strength, the ion starts to hop between the two stable states, which results in amplification of the target signal in the FFT spectra. With a further increase in strength, however, the target signal begins diminishing. These results clearly indicate the basic principle of SR: only with a moderate amount of squeezing signal or Gaussian white noise applied, can we detect the subthreshold periodic signals.

\begin{figure}[htbp]
	\centering
	\includegraphics[width=8.3 cm,height=5.6 cm]{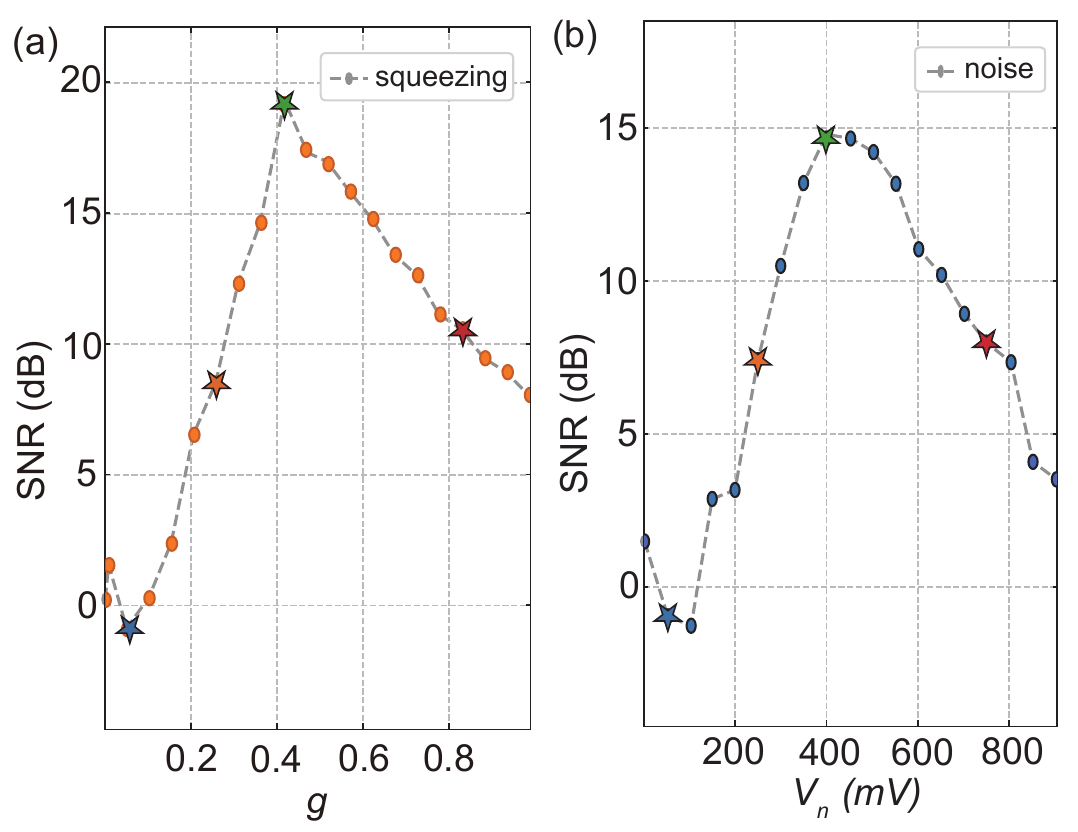}
	\caption{Comparison of SNR between squeezing-induced and noise-induced SRs with respect to the injected squeezing gain and noise strength, where the target signal has a strength of $E_{s}$ = 2357 $\mu$V/cm and a frequency of 1 Hz.
(a) Squeezing-induced SR with a frequency of the squeezing signal of $\omega_{sq}/2\pi$ = 1094.36 kHz. (b) Noise-induced SR with a bandwidth of the Gaussian white noise of $B_{noise}$ =  1 kHz. The lines connect the experimental values to guide the eyes. Each data point is for a 60-s record of the PMT output at a sampling rate of 50 Hz. The stars in different colors indicate the input squeezing gain and noise strength, corresponding to the FFT spectra in the same colors in Fig. \ref{Fig2} (a,b). }
	\label{Fig3}
\end{figure}

To elucidate the mechanism of the squeezing-induced SR and identify the optimal input squeezing gain for signal amplification, we have characterized the SR-based sensor under various squeezing conditions, see Fig. \ref{Fig3}(a) for the SNR versus the input squeezing gain. By fixing the weak-signal amplitude at $E_{s}$ = 2357 $\mu$V/cm, we tune the squeezing gain $g$ from 0 to 0.98. The sampling rate of 50 Hz, well above the weak-signal frequency, ensures the accurate tracking of the frequency hopping events over the average interval of 60 seconds, independent of the bistable state offsets. In particular, the amplification effects at the target signal frequency $\omega_{s}$ can be defined as SNR = 10$\log_{10} [P_{g}(\omega_{s})/\overline{P}_{g}(\omega_{s})$] corresponding to $g$, where $P_{g}(\omega_{s})$ is the Fourier power spectral peak at $\omega_{s}$ and $\overline{P}_{g}(\omega_{s})$ represents the estimated average Fourier power spectrum of the system noise around $\omega_{s}$ (specifically, the latter is accomplished within the range from $\omega_{s}/2\pi$-0.1 Hz to $\omega_{s}/2\pi$+0.1 Hz, excluding the peak at $\omega_{s}/2\pi$=1 Hz). A broad range of squeezing gain enhances the weak signal, with a maximum SNR of 19.23 dB achieved at $g$ = 0.42. This behavior is consistent with the trends in Fig. 2(a,b), confirming a wide operating window for signal amplification via squeezing-induced SR. The location of the maximum in Fig. 3(a) depends on the thermal-noise strength involved experimentally; see details in Appendix A.

Figure \ref{Fig3}(b) presents a reference measurement of the noise-induced SR under the same weak-signal strength as in Fig. \ref{Fig3}(a), where the SNR is plotted as a function of the noise strength $V_{n}$, varying from 0 to 900 mV, and the maximum SNR of 14.77 dB is achieved at $V_{n}$ = 400 mV. Comparing the curves in Fig. \ref{Fig3}(a,b), we see that the maximum SNR obtained via the squeezing-induced SR exceeds that of the noise-induced SR by 4.28 dB, highlighting the advantage of the squeezing-assisted signal detection in electric-field sensing. 

\subsection{Applications in sensing weak signals}
\begin{figure}[htbp]
	\centering
	\includegraphics[width=8.3 cm,height=5.6 cm]{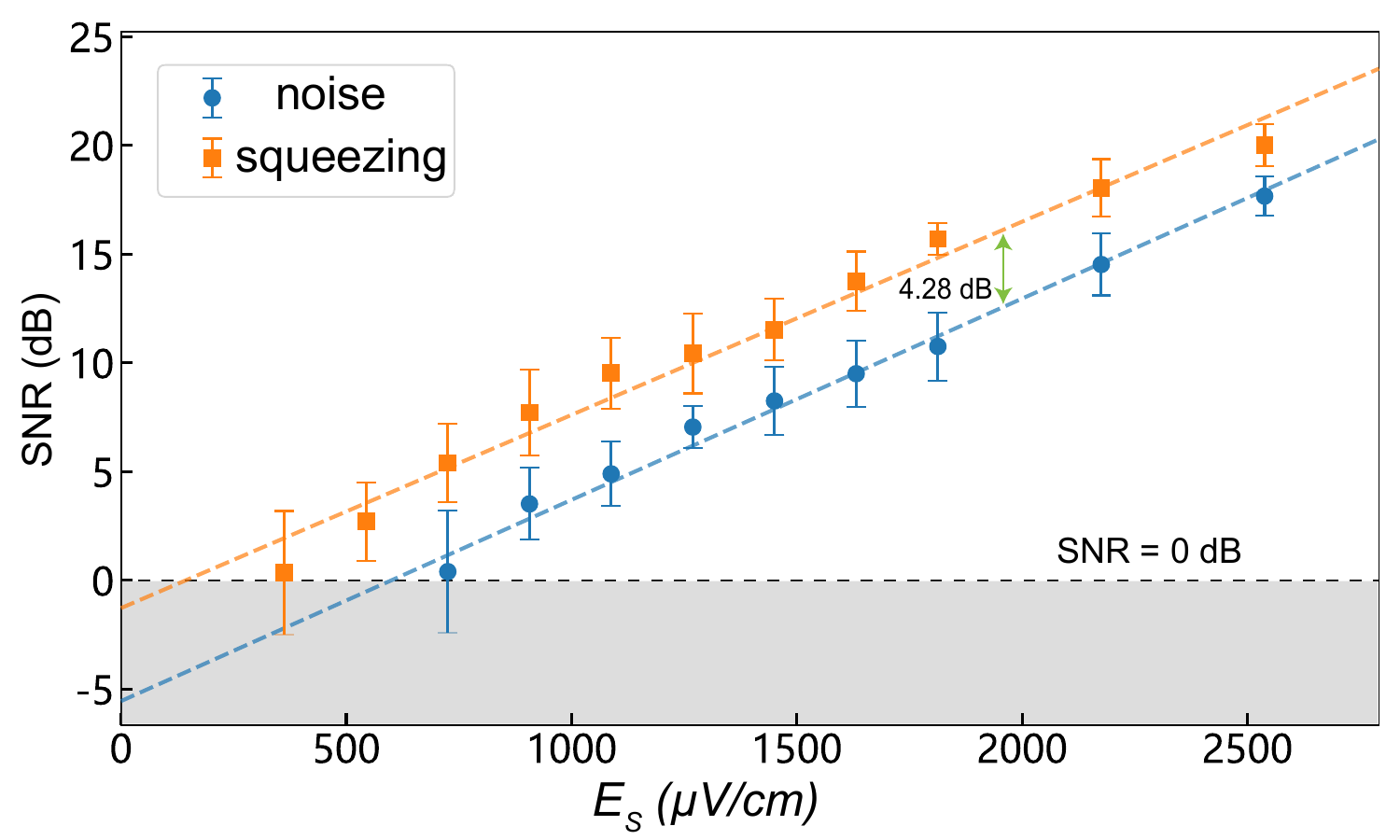}
	\caption{Comparison of SNR between squeezing-induced and noise-induced SRs with respect to the strength of the target signal $E_{s}$, where each value represents the optimal measurement, i.e., the peak value observed in Fig. \ref{Fig3}, obtained by tuning the squeezing gain or noise strength. The dots are experimental data and the dashed lines are linear fits to the experimental data. The error bars indicate the statistical standard deviation of the experimental data obtained from 10 measurements, each with a measurement time of 60 s. The horizontal dashed line corresponds to SNR = 0 dB, i.e., the detection limit; the shaded region indicates the undetectable regime.}
	\label{Fig4}
\end{figure}

Based on the results acquired from Fig. \ref{Fig3}, we have further investigated in Fig. \ref{Fig4} the detection limits of the squeezing-induced SR for weak signals under optimal conditions. Unlike the analysis in Fig. \ref{Fig3} that focuses on the variation of SNR with respect to the squeezing gain and noise strength, here we aim to find the smallest detectable signal when the system is operated with its optimal squeezing gain and noise strength. To this end, we consider weak target signals with a frequency of 1 Hz applied under identical operating conditions, and the SNR is recorded by decreasing the signal strength.
Figure \ref{Fig4} shows the SNR as a function of the target-signal strength $E_{s}$ for both squeezing-induced and noise-induced SRs; the squeezing-induced SR yields an SNR improvement of 4.28 $\pm$ 0.39 dB. The minimum detectable signals, corresponding to 0 dB, exist at $E_{s} = 142 \pm 8~\mu$V/cm and $E_{s} = 617 \pm 34~\mu$V/cm, respectively. We emphasize that the reported minimum detectable signals here depend on the noise present in the system. Acquiring the true detection limit of each method needs further optimization. Nevertheless, considering the fact that noise is inevitable in practical systems, these results underscore the superiority of squeezing-induced SR as an efficient approach to enhancing SNR and improving the measurement precision in stochastic-resonance-based sensing.

\section{Conclusion}
In summary, with a single trapped ion exhibiting Duffing-type nonlinearities, we have achieved squeezing-induced SR for weak-signal amplification, demonstrating a 4.28 $\pm$ 0.39 dB improvement in SNR compared with the conventional noise-induced SR. This new SR approach redistributes the thermal noise in phase space to enable amplification and extraction of weak signals without the need for injecting external noise sources. Experimentally, we have witnessed in the phase space that the hysteresis loop of the oscillator shifts toward a lower driving strength due to amplifying the amplitude noise of the oscillator, which helps reduce the corresponding phase noise and enhances the SNR. Compared with previous trapped-ion electric-field sensing studies \cite{Biercuk2010, Liu2021, Kevin2021, Wei2023}, our experiment is designed for measuring low-frequency electric fields, which is experimentally more challenging and relevant for practical applications. In particular, in contrast with Ref. \cite{ Bonus2025}, who also accomplished electric-field sensing with a low signal frequency, our technique achieved a much better detection sensitivity by amplifying the ion’s vibration and suppressing phase-noise fluctuations. Our technique may find potential applications in underwater low-frequency electric-field detection \cite{Zhang2023ELF, Yu2019UnderwaterEF, Liu2022HighSpeedBoat, Hu2025LOFARBP}, underground geophysical exploration \cite{Lu2022}, geothermal electromagnetic sounding \cite{Zhang2023WEM}, and other related fields \cite{Spichak2009}.

There remains considerable potential for improving the sensitivity of the trapped-ion squeezing-induced SR sensor. For trapped-ion sensors, lowering the ion's heating rate is crucial for enhancing measurement precision. In the present implementation, the hysteretic transition of the Duffing oscillator occurs along the radial direction, i.e., aligned with the rf confinement axis, where rf heating is most pronounced. A possible route toward improvement is to modify the confinement geometry or employ an ion trap with axial asymmetry, allowing the hysteretic response to occur along the axial direction. Such a configuration would significantly reduce rf-induced heating and thereby lead to a substantial enhancement in detection sensitivity.

\section*{Acknowledgments}
This work was supported by National Natural Science Foundation of China under Grants No. 12534020, U25D9005, 12074390, 12304315, 12074346,
by Guangdong Provincial Quantum Science Strategic Initiative under Grants No. GDZX2305004 and GDZX2505001, by Natural Science Foundation of Wuhan under Grant
No. 2024040701010063, by Science and Technology Project of SGCC 52120025005C-338-WLCY, by Nansha Senior Leading Talent Team Project under Grant No. 2021CXTD02, and by the Special Project for Research and Development in Key Areas of Guangdong Province under Grant No. 2020B0303300001.\\

\begin{appendix}
\begin{figure}
	\centering
	\includegraphics[width=8.3 cm,height=6.3 cm]{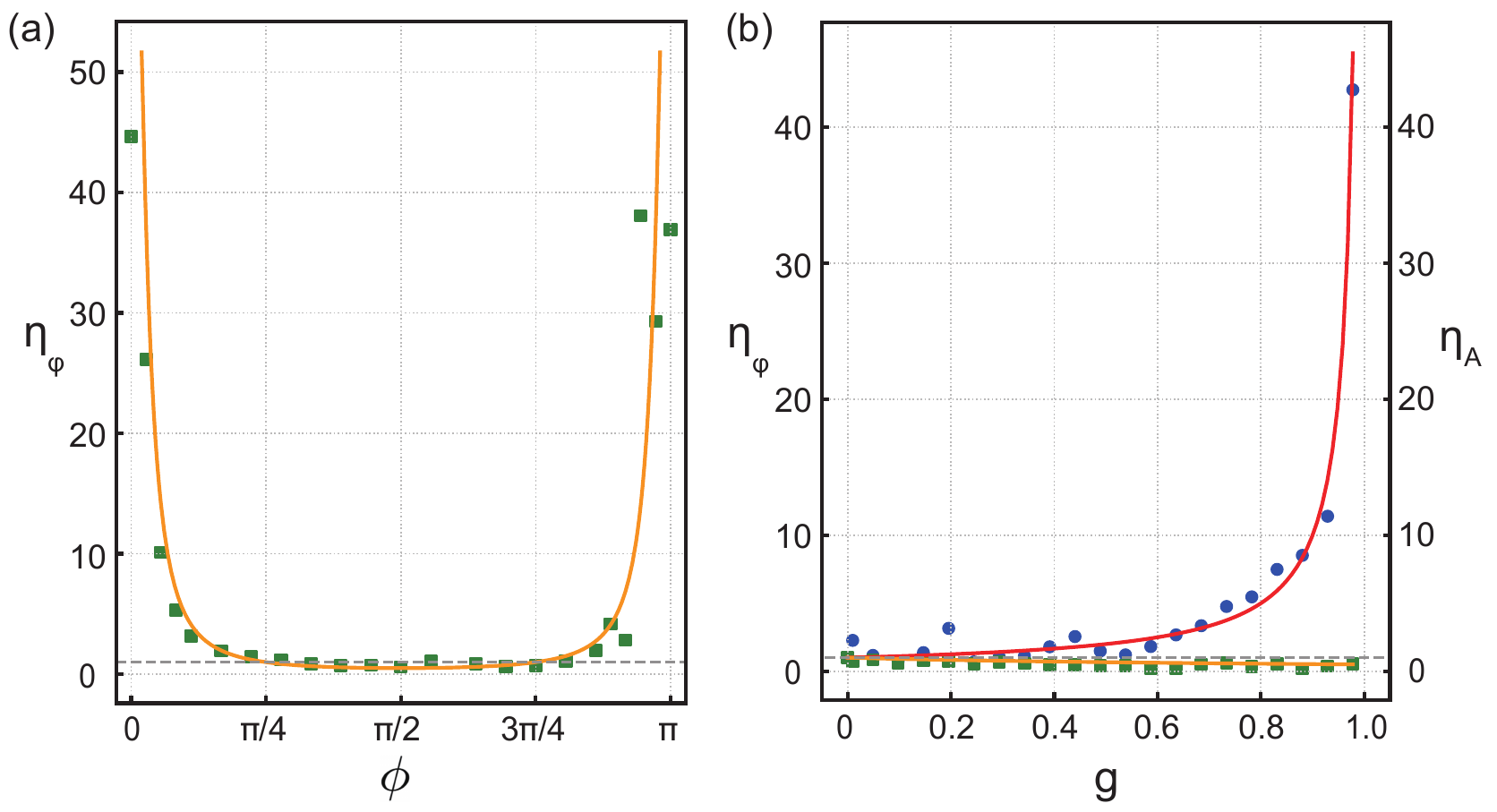}
	\caption{(a) Phase-noise squeezing ratio as a function of  $\phi$, where the squeezing gain is $g$ = 0.98. (b) Amplitude and phase-noise squeezing ratios as functions of $g$, where the squeezing phase is $\phi$ = $\pi/2$. The green and blue dots represent the experimental data for the phase squeezing ratio and amplitude squeezing ratio. Each dot was acquired by 10 measurements. Orange and red lines are the fitted curves from Eq. (\ref{eqa5}) and Eq. (\ref{eqa4}). The dashed line represents the case of no squeezing with $\eta_{A}=\eta_{\phi}=$1.}
	\label{Fig5}
\end{figure}

	\section{Dynamics under Noise}

	Taking into account the influence of thermal noise in amplitude and phase of the Duffing oscillator, i.e., $\delta_{A}(t)$ and $\delta_{\varphi}(t)$, the solution to Eq. (\ref{eq1}) turns out to be $x(t)$ = $[A_{0} + \delta_{A}(t)]\sin [\omega_{x}{t} + \delta_{\varphi}(t) ]$, which can be mathematically rewritten by two orthogonal components $X(t)$ and $Y(t)$ as \cite{Natarajan1995}
\begin{equation}
    x(t) = X(t)\sin(\omega_{x}{t}) + Y(t)\cos(\omega_{x}{t}),
    \label{eqa1}
\end{equation}
    where, in our case, the two orthogonal components can be set as the oscillation amplitude and phase, i.e., $X(t)$ = $A_{0}+\delta_{A}(t)$ and $Y(t)$ =  $A_{0}\delta_{\varphi}(t)$. Then we have the standard deviations of $X$ and $Y$ to be $\sigma_{X}$ = $\sigma_{A}$ and $\sigma_{Y}$ = $A_{0}\sigma_{\varphi}$. When the squeezing is applied to the system, Eq. (\ref{eq1}) is rewritten based on $X(t)$ and $Y(t)$ as
\begin{equation}
    \dot{X}(t)+\frac{\gamma}{2}\left(1+g\cos(2\phi)\right)X(t) = \frac{F_{i} \left[ 1 + q E_{s}  \right] + f_{X}(t)}{2\omega_{x}},
    \label{eqa2}
\end{equation}
\begin{equation}
    \dot{Y}(t)+\frac{\gamma}{2}\left(1-g\cos(2\phi)\right)Y(t) = \frac{ f_{Y}(t)}{2\omega_{x}},
    \label{eqa3}
\end{equation}
where $\gamma$ is the friction coefficient due to the cooling laser and $g$ is the gain related to the modification of the trapping potential with respect to the squeezing signal. For our purposes, we define the stochastic force due to thermal noises as $f_{n}(t) =  f_{X}(t)\cos(\omega_{x}{t}) + f_{Y}(t)\sin(\omega_{x}{t})$. The second-order and fast-oscillating terms at $2\omega_{x}$ are neglected in the above equations \cite{Wei2022}.

From Eq. (\ref{eqa2}) and Eq. (\ref{eqa3}), we see the squeezing effect on the friction coefficients in $X(t)$ and $Y(t)$ in a different fashion. The friction coefficient in  $X(t)$ increases by the same factor, while it decreases in $Y(t)$ by a factor of $g\cos(2\phi)$.  In contrast, in the absence of the squeezing, i.e., when $g = 0$, $f_{X}(t)$ and $f_{Y}(t)$ represent stationary random processes under the thermal noise. In this case, the variance of the temperature-dependent components $X(t)$ and $Y(t)$ is  $\sigma_{X}^{2}(0) = \sigma_{Y}^{2}(0)= \frac{k_{B}T}{2m\omega_{x}^{2}}$ \cite{Majorana1997}. As a result, with squeezing applied, the change in the friction coefficient causes a variation in the secular frequency $\omega_{x}$ of the ion. We acquire the noise squeezing ratios $\eta_{A}$ and $\eta_{\varphi}$ of the amplitude and phase as \cite{Briant2003}
\begin{equation}
    \eta _{A} = \frac{\sigma_{A}^{2}(g,\phi)}{\sigma_{A}^{2}(0)} = \frac{\sigma_{X}^{2}(g,\phi)}{\sigma_{X}^{2}(0)} = \frac{1}{1 + g\cos(2\phi)},
    \label{eqa4}
\end{equation}
\begin{equation}
    \eta _{\varphi} = \frac{\sigma_{\varphi}^{2}(g,\phi)}{\sigma_{\varphi}^{2}(0)} = \frac{\sigma_{Y}^{2}(g,\phi)}{\sigma_{Y}^{2}(0)} = \frac{1}{1 - g\cos(2\phi)},
    \label{eqa5}
\end{equation}
which indicates that the amplitude- and phase-noise squeezing ratios are controlled by the squeezing gain and phase. In our experiment, the phase noise of the oscillator is squeezed. Fig. \ref{Fig5}(a) shows the phase squeezing ratio as a function of $\phi$ with the optimal phase squeezing ratio $\eta_{\varphi}$ approaching 0.5 (i.e., $\phi = \pi/2$). At this phase squeezing intensity, by altering the squeezing gain, Fig. \ref{Fig5}(b) illustrates the measured curves of amplitude and phase squeezing ratios as functions of $g$. The phase and amplitude variances shown in Fig. \ref{Fig5} are obtained using the method described in Appendix B.
From the results, we identify that the best squeezing effect of phase noise that occurs at $g\cos(2\phi) = -1$.

In our experiment, the noise relevant to SR is set by thermal fluctuations of the ion's motional mode at a finite effective temperature. The thermal noise varies depending on the ion's mass, trap frequencies, and cooling conditions, which together determine the effective temperature and the diffusion strength of the motional mode. When the thermal noise is relatively stronger, transitions between the two motional states occur more easily; thus a smaller squeezing gain $g$ can reach the condition of squeezing-induced SR. We consider the SR to be predictable by carefully estimating the characteristic parameters of the system. 

\begin{figure}
	\centering
	\includegraphics[width=8.3 cm,height=6.3 cm]{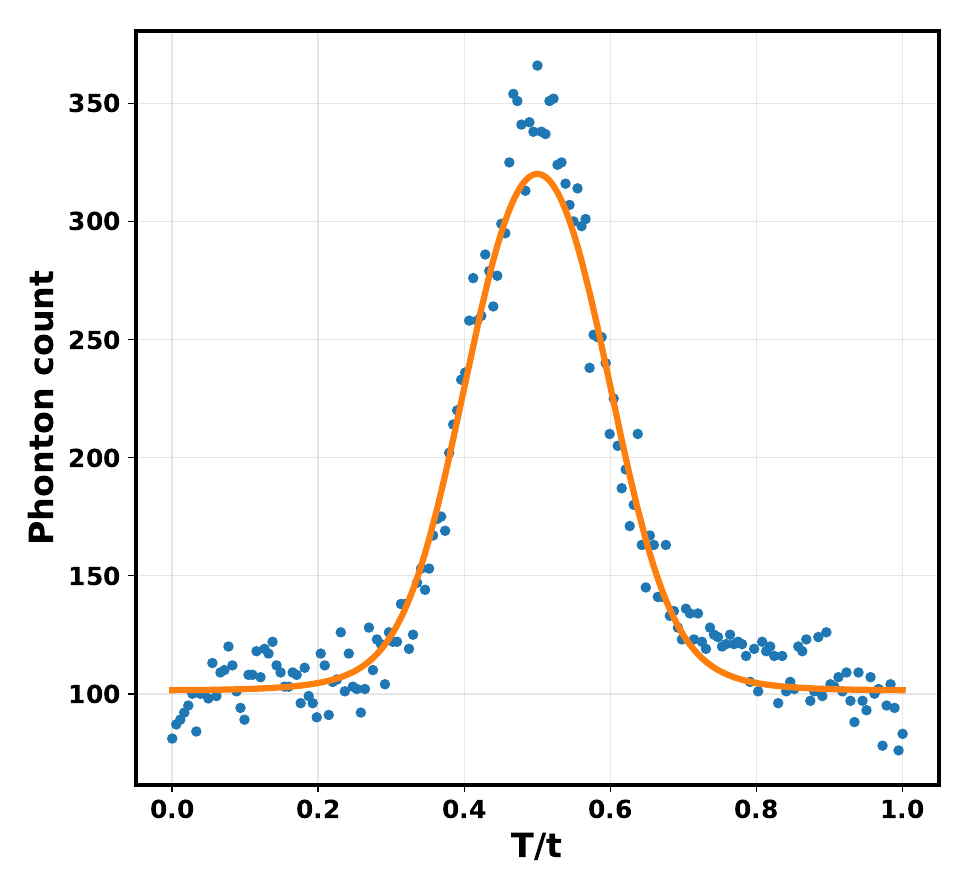}
	\caption{The typical fitting case for accumulated photon counts, from which $A$ = 12.67 $\mu$m and $\varphi$ = 0.017 are acquired under the injection voltage  $V_{i}$ = 40 mV. The dots are experimental data and the solid curves are fitted by Eq. (\ref{eqb2}). Total photon counts $N$ = 2.88$\times$10$^{4}$ are recorded over 183 unit time intervals. Each measurement time is $t_{m}$ = 3 s. We set the parameter values as $\sigma_{t}$ = 60$t_{r}$ = 0.6 $\mu$s with a time-to-amplitude converter resolution $t_{r} = 10 $ ns, $\alpha$ = 3$\times$10$^{-6}$ and $\beta$ = 100. }
	\label{Fig7}
\end{figure}

\section{The Fitting method}
	For our purpose, we have developed an approach to quickly acquire the value of $A$ by fitting the recorded photons \cite{Liu2021}. To this end, we assume that $\varphi$ and $A$ are constants during each measurement time $t_{m}$ by ignoring the short-time noise. Under the irradiation of the 397-nm laser beams, the scattering rate $\rho$ at time $t$ is given by \cite{Leibfried2003}
\begin{equation}
    \rho(t)=\dfrac{{\Gamma}s/(4\pi)}{1 + s + 4\left[ \frac{\Delta - k {\omega} A \cos({\omega}t + \varphi)}{\Gamma}\right]^{2}},
\label{eqb1}
\end{equation}
where $\Gamma$ is the decay rate of $P_{1/2}$, $k$ is the wave vector, $\Delta$ means the laser detuning and $s$ represents the saturation parameter. To fit the experimental photon counts, we have considered the potential noises, such as the projection noise, the laser power noise, etc., for which the Gaussian term $G(t)$ is introduced in fitting the curve of $P(t)$ via convolution,
\begin{equation}
    P(t) = \alpha \rho(t) {\ast} G(t) + \beta,
\label{eqb2}
\end{equation}
where $G(t)$ = $(1/\sqrt{2\pi}\sigma_{t})exp[-t^{2}/(2\sigma_{t}^{2})]$ with $\sigma_{t}$ the degree of time dispersion. $\alpha$ and $\beta$ represent the factors regarding the measurement time $t_{m}$.

Figure \ref{Fig7} presents the fitting to the experimental data under the condition of an injection voltage $V_{i}$ = 40 mV and no squeezing. From the fitting, we acquire the oscillation amplitude $A$ and phase $\varphi$ as constants.

\begin{figure}
	\centering
	\includegraphics[width=8.3 cm,height=6.3 cm]{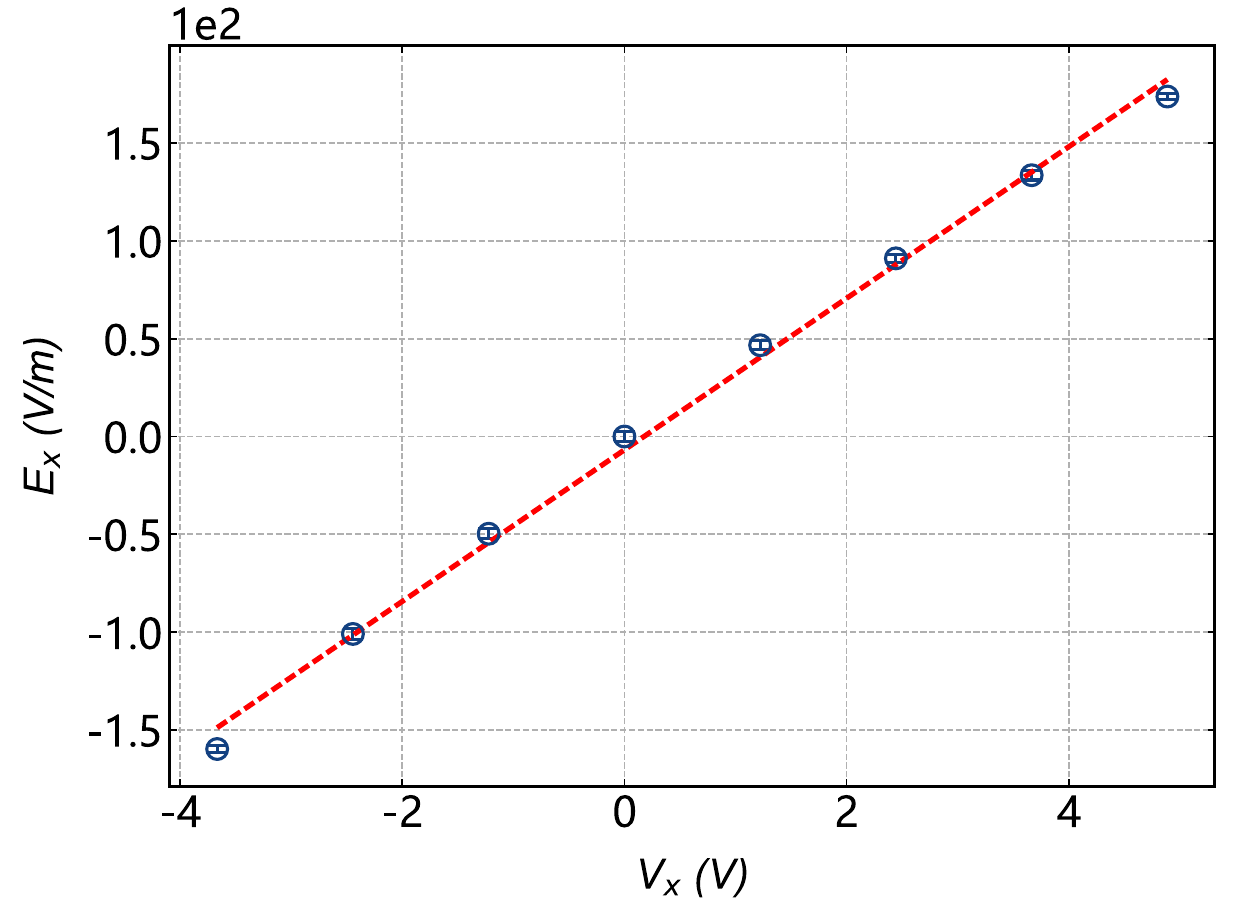}
	\caption{Electric field strength varies with the applied large dc voltage $V_{x}$ for the purpose of calibration. The blue dots represent the experimental data and the dashed line is the linear fitting. The error bars represent the statistical standard deviation of our experimental results based on 10 measurements for each data point.}
	\label{Fig8}
\end{figure}

\section{Calibration for the sensing results}
	For the trapped ion behaving as an oscillator, the electric field to be detected modifies the original trapping potential, an effect that can be described by $\delta q E_{x} = m\omega_{x}^{'2}x' - m\omega_{x}^{2}x$, where $q$ is the electron charge, and both the trap frequency and the ion's position are changed (from $\omega_{x}$ and $x$ to $\omega_{x}^{'}$ and $x^{'}$) due to the additionally applied voltage. For our purposes, we employ a large dc electric field for calibration. To this end, we make measurements of $\partial E_{x}/\partial V_{x}$ by observing the ion’s displacement from the CCD images in the presence and absence of the large dc voltages $V_{x}$ on $SE_{2}$. These measurements are not subject to the electric field driving
    signal, as shown in Fig. \ref{Fig8}, in which ${\partial}E_{x}/{\partial}V_{x} = 38.8 \pm 0.9$ $\mathrm{V\,m^{-1}\,V^{-1}}$ is acquired.
   
 \section{Synchronization} 
   To further strengthen the interpretation of SR, we check the synchronization between switching events and the 1 Hz modulation by defining the offset between the dominant spectral peak and the target modulation frequency $\Delta f=\tilde{f}-1$ Hz,
   where $\tilde{f}$ is the peak frequency extracted from the FFT spectrum. For each squeezing gain $g$, we analyze a single trace using an FFT with a sampling rate of 50 Hz and a record duration of $T = 60$ s (frequency resolution $1 / T = 0.0167$ Hz). When $g$ is small and bistability is absent, the spectrum exhibits no dominant peak near 1 Hz. Once $g$ exceeds the threshold at which bistability emerges, a robust dominant peak appears and remains fixed at $\tilde{f} = 1.016$ Hz for all the measured values of $g$. We find the constant offset $\Delta f = 0.016$ Hz with respect to $g$. This locking and amplification phenomenon near $1$ Hz indicates that the SR is indeed governed by squeezing and noise-activated switching dynamics, rather than by operating-point drift or parametric amplification alone.
    
As such, we acquire the electric-field sensitivity to be $S_E = \sigma_E\sqrt{T} = 1.38 \times 10^3 \mu$V/cm/$\sqrt{Hz}$, where 
the measured electric-field uncertainty is $\sigma_E = 178.1 \mu V/cm $ and the acquisition time T = 60 s per record. 

\end{appendix}

\bibliographystyle{apsrev4-2-title} 
\bibliography{references}

@article{Zhang2023ELF,
  author = {Zhang, Jia-wei and Yu, Peng and Jiang, Run-xiang and Xie, Tao-tao},
  title = {Real-time localization for underwater equipment using an extremely low frequency electric field},
  journal = {Defence Technology},
  volume = {26},
  pages = {203--212},
  year = {2023},
  url = {https://doi.org/10.1016/j.dt.2022.06.014}
}

@article{Yu2019UnderwaterEF,
  author = {Yu, Peng and Cheng, Jinfang and Zhang, Jiawei},
  title = {Ship Target Tracking Using Underwater Electric Field},
  journal = {Prog. Electromagn. Res. M},
  volume = {86},
  pages = {49--57},
  year = {2019},
  url = {https://doi.org/10.2528/PIERM19052001}
}

@article{Liu2022HighSpeedBoat,
  author = {Liu, Qi and Sun, Zhaolong and Jiang, Runxiang and Zhang, Jiawei and Zhu, Kui},
  title = {Electric Field Detection System Based on Denoising Algorithm and High-Speed Motion Platform},
  journal = {Sensors},
  volume = {22},
  number = {14},
  pages = {5118},
  year = {2022},
  url = {https://doi.org/10.3390/s22145118}
}

@article{Hu2025LOFARBP,
  author = {Hu, Huiwen and Sun, Xuepeng and Wang, Guocheng and Liu, Lintao},
  title = {Ocean Target Electric Field Signal Analysis and Detection Using {LOFAR} Based on Basis Pursuit},
  journal = {J. Mar. Sci. Eng.},
  volume = {13},
  number = {2},
  pages = {387},
  year = {2025},
  url = {https://doi.org/10.3390/jmse13020387}
}

@article{Lu2022,
  author = {Lu, Jianxun and Zhuo, Xianjun and Liu, Yong and Zhao, Guoze and Di, Qingyun},
  title = {The Extremely Low Frequency Engineering Project for Underground Exploration},
  journal = {Engineering},
  volume = {10},
  number = {3},
  pages = {13--20},
  year = {2022},
  url = {https://doi.org/10.1016/j.eng.2021.12.003}
}

@article{Zhang2023WEM,
  author = {Zhang, Yilang and Gao, Ya and Fu, Changmin},
  title = {A study of ionospheric impacts in the wireless electromagnetic exploration with the {QWE} method},
  journal = {J. Geophys. Eng.},
  volume = {20},
  number = {2},
  pages = {333--342},
  year = {2023},
  url = {https://doi.org/10.1093/jge/gxad015}
}

@article{Spichak2009,
  author = {Spichak, Viacheslav and Manzella, Adele},
  title = {Electromagnetic sounding of geothermal zones},
  journal = {J. Appl. Geophys.},
  volume = {68},
  number = {4},
  pages = {459--478},
  year = {2009},
  url = {https://doi.org/10.1016/j.jappgeo.2008.05.007}
}

@article{Aslam2023,
  author = {Aslam, Nabeel and Zhou, Hengyun and Urbach, Elana K. and Turner, Matthew J. and Walsworth, Ronald L. and Lukin, Mikhail D. and Park, Hongkun},
  title = {Quantum sensors for biomedical applications},
  journal = {Nat. Rev. Phys.},
  volume = {5},
  pages = {157--169},
  year = {2023},
  url = {https://doi.org/10.1038/s42254-023-00558-3}
}

@article{Yu2021,
  author = {Yu, Chung-Jui and von Kugelgen, Stephen and Laorenza, Daniel W. and Freedman, Danna E.},
  title = {A Molecular Approach to Quantum Sensing},
  journal = {ACS Cent. Sci.},
  volume = {7},
  pages = {712--723},
  year = {2021},
  url = {https://doi.org/10.1021/acscentsci.0c00737}
}

@article{Yu2018,
  author = {Yu, Hang and Martynov, Denis and Vitale, Salvatore and Evans, Matthew and Shoemaker, David and Barr, Bryan and Hammond, Giles and Hild, Stefan and Hough, James and Huttner, Sabina and Rowan, Sheila and Sorazu, Borja and Carbone, Ludovico and Freise, Andreas and Mow-Lowry, Conor and Dooley, Katherine L. and Fulda, Paul and Grote, Hartmut and Sigg, Daniel},
  title = {Prospects for Detecting Gravitational Waves at 5 Hz with Ground-Based Detectors},
  journal = {Phys. Rev. Lett.},
  volume = {120},
  pages = {141102},
  year = {2018},
  url = {https://doi.org/10.1103/PhysRevLett.120.141102}
}

@article{Shao2018,
  author = {Shao, Zhengzheng and Yin, Zhizhen and Song, Helun and Liu, Wei and Li, Xiujian and Zhu, Jubo and Biermann, Klaus and Bonilla, Luis L. and Grahn, Holger T. and Zhang, Yaohui},
  title = {Fast Detection of a Weak Signal by a Stochastic Resonance Induced by a Coherence Resonance in an Excitable GaAs/Al$_{0.45}$Ga$_{0.55}$As Superlattice},
  journal = {Phys. Rev. Lett.},
  volume = {121},
  pages = {086806},
  year = {2018},
  url = {https://doi.org/10.1103/PhysRevLett.121.086806}
}

@article{Badzey2005,
  author = {Badzey, Robert L. and Mohanty, Pritiraj},
  title = {Coherent signal amplification in bistable nanomechanical oscillators by stochastic resonance},
  journal = {Nature},
  volume = {437},
  pages = {995--998},
  year = {2005},
  url = {https://doi.org/10.1038/nature04124}
}

@article{Braakman2014,
  author = {Braakman, F. R. and Cadeddu, D. and Tütüncüoglu, G. and Matteini, F. and Rüffer, D. and Fontcuberta i Morral, A. and Poggio, M.},
  title = {Nonlinear motion and mechanical mixing in as-grown GaAs nanowires},
  journal = {Appl. Phys. Lett.},
  volume = {105},
  pages = {173111},
  year = {2014},
  url = {https://doi.org/10.1063/1.4900928}
}

@article{Dolde2011,
  author = {Dolde, F. and Fedder, H. and Doherty, M. W. and Nöbauer, T. and Rempp, F. and Balasubramanian, G. and Wolf, T. and Reinhard, F. and Hollenberg, L. C. L. and Jelezko, F. and Wrachtrup, J.},
  title = {Electric-field sensing using single diamond spins},
  journal = {Nature Phys.},
  volume = {7},
  pages = {459--463},
  year = {2011},
  url = {https://doi.org/10.1038/nphys1969}
}

@article{Sedlacek2012,
  author = {Sedlacek, Jonathon A. and Schwettmann, Arne and Kübler, Harald and Löw, Robert and Pfau, Tilman and Shaffer, James P.},
  title = {Microwave electrometry with Rydberg atoms in a vapour cell using bright atomic resonances},
  journal = {Nature Phys.},
  volume = {8},
  pages = {819--824},
  year = {2012},
  url = {https://doi.org/10.1038/nphys2423}
}

@article{Jing2020,
  author = {Jing, Mingyong and Hu, Ying and Ma, Jie and Zhang, Hao and Zhang, Linjie and Xiao, Liantuan and Jia, Suotang},
  title = {Atomic superheterodyne receiver based on microwave-dressed Rydberg spectroscopy},
  journal = {Nature Phys.},
  volume = {16},
  pages = {911--915},
  year = {2020},
  url = {https://doi.org/10.1038/s41567-020-0918-5}
}

@article{Holloway2022,
  author = {Holloway, Christopher L. and Prajapati, Nikunjkumar and Artusio-Glimpse, Alexandra B. and Berweger, Samuel and Simons, Matthew T. and Kasahara, Yoshiaki and Alu, Andrea and Ziolkowski, Richard W.},
  title = {Rydberg atom-based field sensing enhancement using a split-ring resonator},
  journal = {Appl. Phys. Lett.},
  volume = {120},
  pages = {204001},
  year = {2022},
  url = {https://doi.org/10.1063/5.0088532}
}

@article{Liu2022,
  author = {Liu, Bang and Zhang, Li-Hua and Liu, Zong-Kai and Zhang, Zheng-Yuan and Zhu, Zhi-Han and Gao, Wei and Guo, Guang-Can and Ding, Dong-Sheng and Shi, Bao-Sen},
  title = {Highly Sensitive Measurement of a Megahertz rf Electric Field with a Rydberg-Atom Sensor},
  journal = {Phys. Rev. Applied},
  volume = {18},
  pages = {014045},
  year = {2022},
  url = {https://doi.org/10.1103/PhysRevApplied.18.014045}
}

@article{Wu2024,
  author = {Wu, Kang-Da and Xie, Chongwu and Li, Chuan-Feng and Guo, Guang-Can and Zou, Chang-Ling and Xiang, Guo-Yong},
  title = {Nonlinearity-enhanced continuous microwave detection based on stochastic resonance},
  journal = {Sci. Adv.},
  volume = {10},
  pages = {eado8130},
  year = {2024},
  url = {https://doi.org/10.1126/sciadv.ado8130}
}

@article{Liu2021,
  author = {Liu, Zhi-chao and Wei, Ya-qi and Chen, Liang and Li, Ji and Dai, Shuang-qing and Zhou, Fei and Feng, Mang},
  title = {Phonon-Laser Ultrasensitive Force Sensor},
  journal = {Phys. Rev. Applied},
  volume = {16},
  pages = {044007},
  year = {2021},
  url = {https://doi.org/10.1103/PhysRevApplied.16.044007}
}

@article{Wei2022,
  author = {Wei, Ya-Qi and Wang, Ying-Zheng and Liu, Zhi-Chao and Cui, Tai-Hao and Chen, Liang and Li, Ji and Dai, Shuang-Qin and Zhou, Fei and Feng, Mang},
  title = {Detection of DC electric forces with zeptonewton sensitivity by single-ion phonon laser},
  journal = {Sci. China-Phys. Mech. Astron.},
  volume = {65},
  pages = {110313},
  year = {2022},
  url = {https://doi.org/10.1007/s11433-022-1954-7}
}

@article{Wei2023,
  author = {Wei, Y.-Q. and Yuan, Q. and Chen, L. and Cui, T.-H. and Li, J. and Dai, S.-Q. and Zhou, F. and Feng, M.},
  title = {Time and Frequency Resolution of Alternating Electric Signals via Single-Atom Sensor},
  journal = {Phys. Rev. Applied},
  volume = {19},
  pages = {064062},
  year = {2023},
  url = {https://doi.org/10.1103/PhysRevApplied.19.064062}
}

@article{Bonus2025,
  author = {Bonus, F. and Knapp, C. and Valahu, C. H. and Mironiuc, M. and Weidt, S. and Hensinger, W. K.},
  title = {Ultrasensitive single-ion electrometry in a magnetic field gradient},
  journal = {Nature Phys.},
  volume = {21},
  pages = {1189--1195},
  year = {2025},
  url = {https://doi.org/10.1038/s41567-025-02887-9}
}

@article{Wu2025,
  author = {Wu, Hao and Mitts, Grant D. and Ho, Clayton Z. C. and Rabinowitz, Joshua A. and Hudson, Eric R.},
  title = {Wideband electric field quantum sensing via motional Raman transitions},
  journal = {Nat. Phys.},
  volume = {21},
  pages = {380--385},
  year = {2025},
  url = {https://doi.org/10.1038/s41567-024-02753-0}
}

@article{Deng2023,
  author = {Deng, Bo and Göb, Moritz and Stickler, Benjamin A. and Masuhr, Max and Singer, Kilian and Wang, Daqing},
  title = {Amplifying a Zeptonewton Force with a Single-Ion Nonlinear Oscillator},
  journal = {Phys. Rev. Lett.},
  volume = {131},
  pages = {153601},
  year = {2023},
  url = {https://doi.org/10.1103/PhysRevLett.131.153601}
}

@article{Biercuk2010,
  author = {Biercuk, Michael J. and Uys, Hermann and Britton, Joe W. and VanDevender, Aaron P. and Bollinger, John J.},
  title = {Ultrasensitive detection of force and displacement using trapped ions},
  journal = {Nat. Nanotech.},
  volume = {5},
  pages = {646--650},
  year = {2010},
  url = {https://doi.org/10.1038/nnano.2010.165}
}

@article{Brownnutt2015,
  author = {Brownnutt, M. and Kumph, M. and Rabl, P. and Blatt, R.},
  title = {Ion-trap measurements of electric-field noise near surfaces},
  journal = {Rev. Mod. Phys.},
  volume = {87},
  pages = {1419},
  year = {2015},
  url = {https://doi.org/10.1103/RevModPhys.87.1419}
}

@book{Gardiner2000,
  author = {Gardiner, C. W. and Zoller, P.},
  title = {Quantum Noise: A Handbook of Markovian and Non-Markovian Quantum Stochastic Methods with Applications to Quantum Optics},
  edition = {2nd},
  publisher = {Springer},
  address = {Berlin},
  year = {2000}
}

@article{Kuo1999,
  author = {Kuo, S. M. and Morgan, D. R.},
  title = {Active noise control: a tutorial review},
  journal = {Proc. IEEE},
  volume = {87},
  pages = {943--973},
  year = {1999},
  url = {https://doi.org/10.1109/5.763310}
}

@article{Khodjasteh2010,
  author = {Khodjasteh, K. and Lidar, D. A. and Viola, L.},
  title = {Arbitrarily accurate dynamical control in open quantum systems},
  journal = {Phys. Rev. Lett.},
  volume = {104},
  pages = {090501},
  year = {2010},
  url = {https://doi.org/10.1103/PhysRevLett.104.090501}
}

@article{Suter2016,
  author = {Suter, D. and Alvarez, G. A.},
  title = {Colloquium: Protecting quantum information against environmental noise},
  journal = {Rev. Mod. Phys.},
  volume = {88},
  pages = {041001},
  year = {2016},
  url = {https://doi.org/10.1103/RevModPhys.88.041001}
}

@article{Gammaitoni1998,
  author = {Gammaitoni, Luca and Hänggi, Peter and Jung, Peter and Marchesoni, Fabio},
  title = {Stochastic resonance},
  journal = {Rev. Mod. Phys.},
  volume = {70},
  pages = {223},
  year = {1998},
  url = {https://doi.org/10.1103/RevModPhys.70.223}
}

@article{Wellens2004,
  author = {Wellens, Thomas and Shatokhin, Vyacheslav and Buchleitner, Andreas},
  title = {Stochastic resonance},
  journal = {Rep. Prog. Phys.},
  volume = {67},
  pages = {45},
  year = {2004},
  url = {https://doi.org/10.1088/0034-4885/67/1/R02}
}

@article{McNamara1988,
  author = {McNamara, Bruce and Wiesenfeld, Kurt and Roy, Rajarshi},
  title = {Observation of Stochastic Resonance in a Ring Laser},
  journal = {Phys. Rev. Lett.},
  volume = {60},
  pages = {2626},
  year = {1988},
  url = {https://doi.org/10.1103/PhysRevLett.60.2626}
}

@article{Hibbs1995,
  author = {Hibbs, A. D. and Singsaas, A. L. and Jacobs, E. W. and Bulsara, A. R. and Bekkedahl, J. J. and Moss, F.},
  title = {Stochastic resonance in a superconducting loop with a Josephson junction},
  journal = {J. Appl. Phys.},
  volume = {77},
  pages = {2582--2590},
  year = {1995},
  url = {https://doi.org/10.1063/1.358720}
}

@article{Wan2014,
  author = {Wan, W. and Wu, H. Y. and Chen, L. and Zhou, F. and Gong, S. J. and Feng, M.},
  title = {Demonstration of motion transduction in a single-ion nonlinear mechanical oscillator},
  journal = {Phys. Rev. A},
  volume = {89},
  pages = {063401},
  year = {2014},
  url = {https://doi.org/10.1103/PhysRevA.89.063401}
}

@article{Natarajan1995,
  author = {Natarajan, Vasant and DiFilippo, Frank and Pritchard, David E.},
  title = {Classical Squeezing of an Oscillator for Subthermal Noise Operation},
  journal = {Phys. Rev. Lett.},
  volume = {74},
  pages = {2855},
  year = {1995},
  url = {https://doi.org/10.1103/PhysRevLett.74.2855}
}

@article{House2008,
  author = {House, M. G.},
  title = {Analytic model for electrostatic fields in surface-electrode ion traps},
  journal = {Phys. Rev. A},
  volume = {78},
  pages = {033402},
  year = {2008},
  url = {https://doi.org/10.1103/PhysRevA.78.033402}
}

@article{Liu2020,
  author = {Liu, Z.-C. and Chen, L. and Li, J. and Zhang, H. and Li, C. and Zhou, F. and Su, S.-L. and Yan, L.-L. and Feng, M.},
  title = {Structural phase transition of the ion crystals embedded in an optical lattice},
  journal = {Phys. Rev. A},
  volume = {102},
  pages = {033116},
  year = {2020},
  url = {https://doi.org/10.1103/PhysRevA.102.033116}
}

@article{Blumel1988,
  author = {Blümel, R. and Chen, J. M. and Peik, E. and Quint, W. and Schleich, W. and Shen, Y. R. and Walther, H.},
  title = {Phase Transitions of Stored Laser-Cooled Ions},
  journal = {Nature},
  volume = {334},
  pages = {309–313},
  year = {1988},
  url = {https://doi.org/10.1038/334309a0}
}

@article{Vahala2009,
  author = {Vahala, K. and Herrmann, M. and Knünz, S. and Batteiger, V. and Saathoff, G. and Hänsch, T. W. and Udem, T.},
  title = {A phonon laser},
  journal = {Nat. Phys.},
  volume = {5},
  pages = {682--686},
  year = {2009},
  url = {https://doi.org/10.1038/nphys1367}
}

@article{Yuan2024,
  author = {Yuan, Q. and Dai, S.-Q. and Li, P.-D. and Wei, Y.-Q. and Li, J. and Zhou, F. and Zhang, J.-Q. and Chen, L. and Feng, M.},
  title = {Stochastic resonance via single-ion phonon laser},
  journal = {Appl. Phys. Lett.},
  volume = {125},
  pages = {102201},
  year = {2024},
  url = {https://doi.org/10.1063/5.0222517}
}

@article{Blums2018,
  author = {Bl{\={u}}ms, Valdis and Piotrowski, Marcin and Hussain, Mahmood I. and Norton, Benjamin G. and Connell, Steven C. and Gensemer, Stephen and Lobino, Mirko and Streed, Erik W.},
  title = {A single-atom {3D} sub-attonewton force sensor},
  journal = {Sci. Adv.},
  volume = {4},
  number = {3},
  pages = {eaao4453},
  year = {2018},
  url = {https://doi.org/10.1126/sciadv.aao4453}
}

@article{Leibfried2003,
  author = {Leibfried, D. and Blatt, R. and Monroe, C. and Wineland, D.},
  title = {Quantum dynamics of single trapped ions},
  journal = {Rev. Mod. Phys.},
  volume = {75},
  pages = {281},
  year = {2003},
  url = {https://doi.org/10.1103/RevModPhys.75.281}
}

@article{Majorana1997,
  author = {Majorana, E. and Ogawa, Y.},
  title = {Mechanical thermal noise in coupled oscillators},
  journal = {Phys. Lett. A},
  volume = {233},
  pages = {162--168},
  year = {1997},
  url = {https://doi.org/10.1016/S0375-9601(97)00458-1}
}

@article{Briant2003,
  author = {Briant, T. and Cohadon, P. F. and Pinard, M. and Heidmann, A.},
  title = {Optical phase-space reconstruction of mirror motion at the attometer level},
  journal = {Eur. Phys. J. D},
  volume = {22},
  pages = {131--140},
  year = {2003},
  url = {https://doi.org/10.1016/S0375-9601(97)00458-1}
}

@article{Maiwald2009,
  author = {Maiwald, Robert and Leibfried, Dietrich and Britton, Joe and Bergquist, James C. and Leuchs, Gerd and Wineland, David J.},
  title = {Stylus ion trap for enhanced access and sensing},
  journal = {Nature Phys},
  volume = {5},
  pages = {551--554},
  year = {2009},
  url = {https://doi.org/10.1038/nphys1311}
}

@article{Kevin2021,
author = {Kevin A. Gilmore  and Matthew Affolter  and Robert J. Lewis-Swan  and Diego Barberena  and Elena Jordan  and Ana Maria Rey  and John J. Bollinger },
title = {Quantum-enhanced sensing of displacements and electric fields with two-dimensional trapped-ion crystals},
journal = {Science},
volume = {373},
pages = {673-678},
year = {2021},
URL = {https://www.science.org/doi/abs/10.1126/science.abi5226}
}

\end{document}